\documentclass[12pt, letterpaper]{article}

\usepackage{amsmath, amsfonts, amssymb, latexsym, graphics, setspace, graphicx, lineno, sectsty, fullpage, natbib}

\begin{document}

\begin{titlepage}
	\begin{center}
		\large 
		\textsc {Computability, G\"odel's Incompleteness Theorem, and an Inherent Limit on the Predictability of Evolution}
		\vspace{15mm}

		\normalsize Troy Day$^{1,2}$
		\vspace{5mm}
		
		1. Department of Mathematics and Statistics,\\
		Jeffery Hall, Queen's University,\\
		Kingston, ON, K7L 3N6, Canada\\
		tday@mast.queensu.ca, Ph: 613-533-2431, Fax: 613-533-2964
		\vspace{5mm}

		2. Department of Biology, Queen's University\\
		Kingston, ON, K7L 3N6, Canada

	\end{center}
	\vspace{15mm}
		
	\today
\end{titlepage}

\begin{titlepage}

\subsubsection*{Abstract}
\noindent The process of evolutionary diversification unfolds in a vast genotypic space of potential outcomes. During the past century there have been remarkable advances in the development of theory for this diversification \citep{Fisher:1930, Wright:1984, Hofbauer:1988, Lynch:1998, Burger:2000, Ewens:2004, Barton:2007}, and the theory's success rests, in part, on the scope of its applicability. A great deal of this theory focuses on a relatively small subset of the space of potential genotypes, chosen largely based on historical or contemporary patterns, and then predicts the evolutionary dynamics within this pre-defined set. To what extent can such an approach be pushed to a broader perspective that accounts for the potential open-endedness of evolutionary diversification? There have been a number of significant theoretical developments along these lines \citep{Gillespie:1984, Fontana:1994, Szathmary:1995, MaynardSmith:1995, Wagner:1996, Orr:1998, Stadler:2001, Yedid:2002, Orr:2002, Wagner:2003, Fernando:2007, Nowak:2008, Joyce:2008, Ohtsuki:2009, Manapat:2009} but the question of how far such theory can be pushed has not been addressed. Here a theorem is proven demonstrating that, because of the digital nature of inheritance, there are inherent limits on the kinds of questions that can be answered using such an approach. In particular, even in extremely simple evolutionary systems a complete theory accounting for the potential open-endedness of evolution is unattainable unless evolution is progressive. The theorem is closely related to G\"odel's Incompleteness Theorem \citep{Godel:1931, Nagel:1958, Davis:1965, Heijenoort:1967} and to the Halting Problem from computability theory \citep{Turing:1936, Cutland:1980}.

\end{titlepage}

\linenumbers
\doublespace

\raggedright

\section*{Introduction}

Much of evolutionary theory is, in an important sense, fundamentally historical. The process of evolutionary diversification unfolds in a vast genotypic space of potential outcomes, and explores some parts of this space and not others. Nevertheless, a great deal of current theory restricts attention to a relatively small subset of this space, chosen largely based on historical or contemporary patterns, and then predicts evolutionary dynamics. Although this can work well for making short-term predictions, ultimately it must fail once evolution gives rise to genuinely novel genotypes lying outside this predefined set \citep{Yedid:2002}. 

This potential limitation on the predictive ability of many models of evolution has been noted on various occasions throughout the development of evolutionary theory \citep{Levinton:1988, Fontana:1994, Wagner:1996, Yedid:2002}, perhaps most famously by Dutch biologist Hugo DeVries when he remarked that ``Natural selection may explain the survival of the fittest, but it cannot explain the arrival of the fittest" \citep{DeVries:1904}. Such statements hint at the notion that many models of evolution are what we might call `local', or `closed', in the sense that they focus attention on a very small (local) region of the evolutionary tree and do not account for the possibility that evolution is an open-ended process. 

The distinction between `closed' and `open-ended' models of evolution will be discussed in more detail below, but in recent years there have been several interesting studies published that are beginning to push the boundaries of analyses towards what we might naturally call open-ended models. These studies include models of abstract replicator populations \citep{Fontana:1994, Szathmary:1995, Nowak:2008, Ohtsuki:2009, Manapat:2009}, models exploring the space of evolutionary possibilities \citep{Fontana:1998b, Stadler:2001, Wagner:2003}, analyses of evolutionary transitions \citep{MaynardSmith:1995, Fontana:1998a}, models for predicting the distribution of allelic effects during evolution \citep{Gillespie:1984, Orr:1998, Orr:2002, Joyce:2008}, and studies of evolvability \citep{Wagner:1996}. Similarly, there have also been many \textit{in silico} and artificial life experiments that explore generic, emergent, properties of evolution \citep{Fontana:1994, Lenski:1999, Yedid:2001, Wilke:2001, Yedid:2002, Lenski:2003, Chow:2004, Ostrowski:2007, Fernando:2007, Yedid:2008, Yedid:2009}. In general these analyses have demonstrated that, once we allow for more open-ended evolution, a much richer suite of evolutionary possibilities arises.

The above studies collectively suggest that accounting for open-ended evolution in theory can yield interesting new insights, and it can also yield new testable predictions \citep{Gillespie:1984, Orr:1998, Orr:2002, Joyce:2008}. Nevertheless, there is still a relative paucity of theoretical studies that allow for open-ended evolution, and so we might expect that much is yet to be learned by broadening evolutionary theory further in this way. My purpose with this article is therefore twofold. First, I simply wish to highlight the fact that there is an important distinction to be made between open-ended versus closed models of evolution (defined more precisely below), and to suggest that open-ended models might more faithfully represent the evolutionary process. Second, and more significantly, I wish to consider whether a push towards a predictive theory that embraces the potential open-endedness of evolution is likely to face additional obstacles, over and above those faced by closed models of evolution. Put another way, I ask the question: To what extent is the development of a predictive, open-ended evolutionary theory possible? 

Although a complete answer to the above question is not possible, in what follows I will provide at least a partial answer. Furthermore, I demonstrate that this answer has interesting connections to the Halting Problem from computability theory and to  G\"odel's Incompleteness Theorem from mathematical logic. In particular, I will use results from these areas to prove a theorem that formally links the concept of progressive evolution to the possibility of developing such a predictive open-ended theory. There remains debate over if, and when, evolution might be progressive \citep{Dawkins:1997, Gould:1997, Adami:2000} and part of this debate stems from the lack of a precise yet general definition of progression. Thus, another way to view the results presented here is as providing such a definition. I will return to this point more fully in the discussion.

\section*{A Motivating Example}

To sharpen the focus on these somewhat abstract ideas, it is worth beginning with a concrete motivating example involving evolutionary prediction. This section does so, focussing primarily on the broad conceptual issues involved. The section that follows then addresses these issues more precisely. 

Consider trying to use evolutionary theory to predict the dynamics of human influenza. Specifically, consider trying to answer the following question: is it likely that a pandemic with the 1918 Spanish influenza strain will ever occur again? This is obviously a difficult, and still somewhat loosely defined, question so let's narrow things down further. One reason we might be skeptical about our ability to make such predictions is because of uncertainty in initial conditions and parameter values, as well as uncertainty about the evolutionary processes involved. In other words, perhaps we lack all of the information required to make such predictions. Furthermore, unexpected contingencies might thwart what would otherwise be accurate predictions. For example, an unanticipated volcanic eruption might temporarily alter commercial air travel patterns, and this might thereby alter the epidemiological and evolutionary dynamics of influenza. 

These practical limitations are clearly important, but are they the only obstacle to making accurate evolutionary predictions or are there other, `inherent', limitations as well. Does the difficulty of making evolutionary predictions stem simply from our lack of knowledge of the evolutionary processes involved or are there reasons why, even in principle, such evolutionary predictions are not possible? 

It is this latter question that is the focus of this article, and therefore I will, at least temporarily, put the above practical concerns aside. Specifically, let's assume that we can build a model that adequately captures all of the relevant evolutionary processes, and that we can obtain all parameter estimates necessary to use such a model. Without getting too much into the specifics, one of the first things we would need to decide is the relevant strain space for the model. The simplest scenario would consider only two strains (e.g., the 1918 strain and the current, predominant, strain). More sophisticated scenarios might instead include several strains that are thought to be important in the dynamics. In either case, both such resulting models would be `closed' in the sense described in the introduction because they focus only on a finite (and relatively small) number of strains. Furthermore, given that there is a discrete and finite number of people who can be infected at any given time, there is then also a finite (and relatively small) number of possible evolutionary outcomes. As will be detailed more precisely later, this then implies that the process will either reach a steady state or it will display periodic behaviour (see Appendix \ref{stoch}). Hence, if a closed model is an accurate description of the evolutionary process, then in principle we can answer the above question by simply running the model until one of these two outcomes occurs. At that point we need only observe whether or not a 1918 Spanish flu pandemic ever occurred during the run of the model (or if it occurred with significant probability). 

But what if the evolutionary process is, instead, open-ended? To explore this possibility we need to be more specific about what is meant by open-ended. Consider again the influenza example. Influenza A has a genome size of more that 12,000 nucleotides, and therefore the number of possible genotypes is enormous. To gain some perspective on just how many genotypes are possible, let's restrict attention to only the smallest of the eight genomic segments of influenza. In this case there are then only approximately 800 nucleotides and therefore approximately $4^{800}$ different possible genotypes. To put this number in perspective, it is approximately $10^{400}$ times larger than the estimated number of atoms in the universe. For a model to be open-ended it would have to allow for such a vast set of possible evolutionary outcomes so that, as in reality, evolutionary change could continue unabated, producing potentially novel outcomes essentially indefinitely. The simplest way we might try to capture this theroetically is to assume that the space of possible genotypes is infinite. 


Given these considerations, if evolutionary theory is to capture an open-ended evolutionary process, then its state space must be effectively infinite. This is necessary but it is not a sufficient condition for open-ended evolution. For example, many stochastic Markovian models in population genetics have an infinite state space (e.g., the infinite alleles model; \cite{Kimura:1964}) but nevertheless do not display open-ended evolution. Rather, further assumptions are often made, such as the assumption that the Markov chain is irreducible and positively recurrent. These assumptions are usually made primarily for mathematical convenience but they rule out the possibility of open-ended evolution since they then guarantee the existence a single unique equilibrium or stationary distribution. As a result, such models cannot capture the possibility that evolutionary change might continue indefinitely. 

What if we relax these assumptions and allow for truly open-ended evolution in the theory that we develop? Are there then even further problems associated with making evolutionary predictions? For example, does this make answering the question about influenza evolution laid out at the start of this section more difficult? You might suspect that the answer is `yes'; at least, the approach suggested above for closed models will no longer suffice because the evolutionary process is no longer guaranteed to settle down to an equilibrium or stationary distribution. Thus, the best we can possibly hope for is that there is some way to prove, using the structure of the model, whether or not such an outcome will occur. Thus, all practical difficulties of predicting evolution aside, it is not obvious whether we can answer the above sort of question about influenza evolution, even in principle.

These issues are now starting to tread heavily into the fields of computability and mathematical logic and, roughly speaking, a theory that can answer the above kind of question about influenza evolution is referred to as a negation-complete theory. This terminology reflects the idea that the theory is complete in the sense of one being able to determine whether a given statement is true, or whether its formal negation is true instead. For example, in the context of influenza, a negation-complete theory would be able to predict whether the statement `the Spanish flu will happen again' is true or whether its formal negation `it is not true that the Spanish flu will happen again' is true instead. More generally, a negation-complete evolutionary theory would be one from which we could determine those parts of genotypic space will be explored by evolution and those that will not. 

Is such a negation-complete theory possible once we allow for open-ended evolution? In the remainder of this article I show that the answer to this question is closely related to the idea of progressive evolution. In particular, even if the system of evolution were simple enough for us to understand everything about how its genetic composition changes from one generation to the next, the following theorem is proven:

\emph{Theorem: A negation-complete evolutionary theory is possible if, and only if, the evolutionary process is progressive.}

The above theorem will be made more precise shortly, but as already alluded to above, it stems from the fact that DNA affords evolution a mechanism of digital inheritance.  As Maynard Smith and Szathm\'ary have noted \citep{MaynardSmith:1995} the combinatorial complexity that arises thereby allows evolution to be effectively open-ended. Indeed, as will be argued below, digital inheritance allows one to characterize evolution (i.e., the change in genetic composition of a population) as a dynamical system on the natural numbers, and therefore the theorem proved below holds for any such dynamical system, not just those meant to model evolution. As a result, the theorem is closely related to other results from mathematics and computer science; namely G\"odel's Incompleteness Theorem \citep{Godel:1931, Nagel:1958, Davis:1965, Heijenoort:1967} and to the Halting Problem from computability theory \citep{Turing:1936, Cutland:1980}. 

\section*{Statement and Proof of Theorem}

In order to give precision to the above theorem, we must specify what is meant by `the evolutionary process', as well as what it means for evolutionary theory to be negation-complete.  The goal is to determine if, even in extremely simple evolutionary processes, there is some inherent limitation on evolutionary theory.

To this end, consider a simplified evolutionary process in which there is a well-mixed population of replicators with some maximal population size, and in which each replicator contains a single piece of DNA. This genetic code can mutate in both composition, and in length, with no pre-imposed bounds. Suppose that each replicator survives and reproduces in a way that depends only on the current genetic composition of the population. For additional simplicity, suppose that generations are discrete. All conclusions hold if events occur in continuous time instead (Appendix \ref{stoch}). Finally, for simplicity of exposition, I will usually assume that the evolutionary dynamics are deterministic in the main text. Again, all results generalize to the case of stochastic evolutionary dynamics, albeit with a few additional assumptions (Appendix \ref{stoch}).

With the above evolutionary dynamic, the genetic composition of the system will evolve over time, and we can characterize the state of the system at any time by the number of each type of replicator (e.g., the number of infections with each possible genotype of influenza). The goal then is to determine if it is possible to construct an evolutionary theory that can predict which parts of the space of potential evolutionary outcomes will be explored during evolutionary diversification, and which will not. Formally, the results presented below are valid for any theory whose derived statements are recursively enumerable. Axiomatic theories are one such example but (roughly speaking) any theoretical approach that can, in principle,  be implemented by a computer falls into this category (Appendix \ref{theory}). Indeed, the statement and proof of the theorem relies on several ideas from computability theory (Appendix \ref{comp}).


The digital nature of inheritance provided by DNA means that, in principle, the number of distinct kinds of replicators that are possible is discrete and unbounded, a property Maynard Smith and Szathm\'ary refer to as `indefinite' heredity \citep{MaynardSmith:1995}. It is indefinite heredity that allows for open-ended evolution. As a result, in principle, the set of possible population states during evolution is isomorphic to the positive integers;  i.e., there exists a one-to-one correspondence between the set of possible population states and the positive integers. Such sets are called denumerable, and in fact the set of population states is effectively denumerable in a computability sense (Appendix \ref{popstate}). Thus we can effectively assign a unique integer-valued `code' to every possible population state. 

In practice, of course, there are limits on the number of kinds of replicators possible, if only because of a finite pool of the required chemical building blocks. Nevertheless, as mentioned earlier the combinatorial nature of indefinite heredity means that the actual number of possible population states is so large as to be effectively infinite.  For simplicity of exposition, it is assumed in the main text that the set of possible population states is truly infinite; however, Appendix \ref{Infinity} makes the notion of `effectively infinite' precise and provides the analogous results for this case.

With the above coding we can formalize evolution mathematically as a mapping of the positive integers to themselves. For example, in the deterministic case we might start with a model (e.g., a mapping $F$) that tells us the number of individuals of each genotype in the next time step, as a function of the current numbers. Then, under the above coding, if $E(n)$ denotes the population state (formally, its integer code number) at time $n$, the model can be recast as a single-variable, integer, mapping $E(n+1)=G(E(n))$ for some function $G$, along with some initial condition. Similarly, in the stochastic case, if we start with a probabilistic mapping $F$, then it can be recast as a mapping $E(n+1)=H(E(n))$ where $H$ gives the probability distribution over the set of code numbers in the next time step as a function of its current distribution (and $E$ is then a vector of probabilities over the integers). Therefore, in general, we can view the evolutionary trajectory as being simply an integer-valued function with an integer-valued argument. Of course, different ways of coding the population states will correspond to different maps, $G$ or $H$, and thus different functions $E(n)$. Also note that the domain of $G$ or $H$ need not be all of the positive integers, and in fact different initial conditions might give rise to different domains as well. This would correspond to there being different basins of attraction in the evolutionary process.

It is also worth noting that, although we have assumed the evolutionary mapping (i.e., $G$ or $H$) is a function of the current genetic composition of the population only, we can relax this assumption and allow evolutionary change to depend on other aspects of the environment as well. In particular, we might expand our definition of `population state' to include both genetic state, and the state of other variables associated with the environment in which the genes exist. Again, as long as such generalized processes can be recast as dynamical systems on the natural numbers, all of the results presented here continue to hold.

The above arguments illustrate how we can view evolution as a dynamical system on the natural numbers, and they also now allow us to formalize the notion of open-ended evolution. In the deterministic setting evolution is open-ended if the mapping $G$ never revisits a previously visited state. Likewise, in the stochastic setting, evolution is open-ended if the mapping $H$ always admits at least one new state each generation with positive probability.

Because we can view evolution as a dynamical system on the natural numbers, evolutionary theory can be viewed as a set of specific rules for manipulating and deducing statements about such numbers. Computability theory deals with functions that map positive integers to themselves, and thus provides a natural set of tools to analyze the problem. A function is called `computable' if there exists some algorithmic procedure that can be followed to evaluate the function in a finite number of steps (Appendix \ref{comp}). 

Again, focusing on the deterministic case, given the assumption that we are able to predict the state of the population from one time step to the next, the function $E(n)$ is computable (see Appendix \ref{comp}). Furthermore, the set of all computable functions is denumerable \citep{Cutland:1980}. Therefore, denoting the $k^{th}$ such function by $\phi_k(n)$, it is clear the evolutionary process, $E(n)$, must correspond to a member of this set. Denote this specific member by $\phi_E(n)$, and again note that, if we change the integer-coding used to identify specific population states, we will obtain a different function $\hat{E}(n)$, and thus a different member of the set,  $\phi_{\hat{E}}(n)$ (Fig. 1).

During evolution, a set of population states will be visited over time (in the stochastic case we consider a state as being visited if the probability of it occurring at some point is larger than a threshold value; Appendix \ref{stoch}). These will be referred to as `evolutionarily attainable' states. In terms of our formalism, this corresponds to the function $\phi_E(n)$ taking on various values of its range, $R_E$, as $n$ increases (Fig. 1). A negation-complete evolutionary theory would be one that can determine whether a code, $x$, satisfies $x \in R_E$ or whether it satisfies $x \notin R_E$ instead. In the language of computability theory, this corresponds to asking whether the predicate `$x \in R_E$' is decidable (Appendix \ref{comp}; \citep{Cutland:1980}). In terms of the influenza example presented earlier, if $x$ is the population state corresponding to a pandemic with the 1918 strain, then the statement `the Spanish flu will happen again' corresponds to the number-theoretic statement $x \in R_E$. Likewise, the statement `it is not true that the Spanish flu will happen again' corresponds to the number-theoretic statement $x \notin R_E$.

Lastly, we can give a precise definition of progressive evolution. Intuitively, evolution is progressive if there is some quantifiable characteristic of the population that increases through evolutionary time. In terms of the above formalization, this means there is a way to recode the population states such that the code number increases during evolution. Formally, evolution is progressive if there exists a computable, one-to-one, coding of the population states by positive integers, $\hat{C}$, such that the corresponding description of the evolutionary process, $\phi_{\hat{E}}(n)$, satisfies $\phi_{\hat{E}}(n+1) > \phi_{\hat{E}}(n)$ for all $n$. Again, in terms of the influenza example presented earlier, if evolution were progressive, then there would be some way to \textit{a priori} code the population states such that, as influenza evolution occurs, the code number of the population increases (I will return to this definition of progression in more detail in the discussion).

We can now rephrase the theorem in terms of precise, technical, language:

\emph{Theorem: `$x \in R_E$' is decidable if, and only if, there exists a computable, one-to-one, coding of the population states by positive integers, $\hat{C}$, such that the corresponding description of the evolutionary process, $\phi_{\hat{E}}(n)$, satisfies $\phi_{\hat{E}}(n+1) > \phi_{\hat{E}}(n)$ for all $n$.}

Proof (Figure 1; see Appendices \ref{comp} and \ref{re} for additional details):

Part 1: If there exists a coding $\hat{C}$ such that $\phi_{\hat{E}}(n+1) > \phi_{\hat{E}}(n) \mbox{ }$ for all $n$ then the predicate `$x \in R_E$' is decidable.

By hypothesis there exists a computable bijection $\hat{C}$ such that, for the corresponding description of the evolutionary process, $\phi_{\hat{E}}(n+1) > \phi_{\hat{E}}(n)$ for all $n$. For any population state, $x$, in the original coding, let $\hat{x}$ be the corresponding code under the bijection $\hat{C}$, and define $z(\hat{x})=\mu i (\phi_{\hat{E}}(i) \geq \hat{x})$, where $\mu i (H(i))$ denotes the minimum value of $i$ for which the argument $H(i)$ is true (Appendix \ref{comp}). Further, define $R_k(n)=\{x:\phi_k(i)=x, i \leq n \}$ (i.e., the range of $\phi_k(n)$ visited by step $n$; Appendix \ref{comp}). Clearly `$\hat{x} \in R_{\hat{E}}(z(\hat{x}))$' is decidable since $ R_{\hat{E}}(z(\hat{x}))$ is finite and can be enumerated, and furthermore $\hat{x} \in R_{\hat{E}}(z(\hat{x})) \Leftrightarrow \hat{x} \in R_{\hat{E}}$ owing to the progressive nature of evolution. Therefore, `$\hat{x} \in R_{\hat{E}}$' is decidable as well. Finally, using $S$ denote the set of population states that are evolutionarily attainable, we have that $\hat{x} \in R_{\hat{E}} \Leftrightarrow \hat{C}^{-1} \hat{x} \in S \Leftrightarrow C\hat{C}^{-1} \hat{x} \in R_E$. Noting that, by definition, $x=C\hat{C}^{-1} \hat{x}$, we obtain $\hat{x} \in R_{\hat{E}} \Leftrightarrow x \in R_E$. Thus, `$ x \in R_E$' is decidable as well.

Part 2: If the predicate `$x \in R_E$' is decidable then there exists a coding $\hat{C}$ such that $\phi_{\hat{E}}(n+1) > \phi_{\hat{E}}(n)$ for all $n$.

We can construct the required computable bijection between population states and an appropriate coding as follows. First, take any effective coding of population states. By hypothesis `$x \in R_E$' is decidable and therefore we can proceed through the population states, $x$, in increasing order, applying the following algorithm:

(i) if $x \notin R_E$ and it is the $k^{th}$ such state up to that point, use the $k^{th}$ odd number as its new code.

(ii) if $x \in R_E$, calculate $\mu i (\phi_E(i)=x)$, and use the $i^{th}$ even number as its new code.

Thus, $R_{\hat{E}}$ is the set of even numbers, and they are visited in increasing order as evolution proceeds. In particular, using $\hat{C} C^{-1}$ to denote the above mapping described in points (i) and (ii), where $C^{-1}$ is the inverse mapping of the coding that generated $x$ (i.e., it takes code $x$ and returns the corresponding population state, $s$), we have $\phi_{\hat{E}}(n+1) = \hat{C} C^{-1}\phi_{E}(n+1) = 2(n+1)$. The last equality follows from the fact that $\hat{C} C^{-1}\phi_{E}(n+1)$ determines the time at which state $\phi_{E}(n+1)$ occurs (which is $n+1$), and assigns it a new code equal to twice this value (point (ii) above). Therefore $\phi_{\hat{E}}(n+1) > \phi_{\hat{E}}(n) \mbox{ } \forall n$. 

Q.E.D.


\section*{Discussion}

This article has two main goals. The first goal is to highlight the distinction between open-ended versus closed models of evolution, and to suggest that open-ended models might better capture real evolutionary processes. The second goal is to explore the extent to which the development of a predictive, open-ended theory of evolution is possible. The above theorem illustrates that there is an interesting connection between this question and analyses from computability theory and mathematical logic. It also draws a formal connection between the extent to which such a theory is possible and the notion of progressive evolution. 

Because the theorem states an equivalence relationship between the possibility of developing a negation-complete theory and progressive evolution, it can be read in two distinct ways. First, it states that if evolution is progressive then a negation-complete theory is possible. This is, perhaps, not too surprising. If evolution is progressive then there would be a good deal of regularity to the process that one ought to be able to exploit in constructing theory. The second way to read the theorem is from the perspective of the reverse implication. This is somewhat more surprising; it states that if evolution is not progressive then a negation-complete theory will not be possible.

These results rest on the fact that digital inheritance allows evolution to be open-ended \citep{MaynardSmith:1995}. If, instead, the hereditary system allowed for only a finite number of discrete possible types, then evolution would either display periodic behaviour or would reach an equilibrium (possibly with stochastic fluctuations; Appendix \ref{stoch}). A negation-complete theory of evolution would then be trivially possible in such cases because, in principle, we could simply develop a finite list of all evolutionary outcomes that can occur (as described in the influenza example earlier).

Of course, despite the existence of digital inheritance, there is nevertheless presumably a bound on the number of population states possible for a variety of reasons. Even so, however, the combinatorial nature of digital inheritance means that the number of possible population states might be considered effectively infinite. An analogous theorem can be proven in such cases by replacing the notion of infinite with a precise notion of effectively infinite instead (Appendix \ref{Infinity}). Likewise, although the main results of the text assume that evolution is deterministic, an analogous theorem holds that accounts for the inherently stochastic nature of the evolutionary process (Appendix \ref{stoch}).

The notion of progressive evolution is somewhat slippery, and there does not exist a general yet precise definition of progression that is universally agreed upon. As a result, this has led to disagreement over the extent to which progressive evolution occurs \citep{Dawkins:1997, Gould:1997}. A complete discussion of the idea of progressive evolution is beyond the scope of this article but a few points are worth making here. 

Most discussions of progressive evolution involve quantities like mean fitness, body size, complexity, or other relatively conspicuous biological measurements. Many such discussions also are retrospective in the sense that they look at historical patterns when attempting to find patterns of progression. But both of these aspects of discussions of progression are problematic. First, although it would be nice to readily identify some obvious, and biologically meaningful, characteristic of a population that changes in a directional way, there is no reason to expect that we have currently thought of all the possibilities. Thus, when defining progression, it would seem desirable to do so in a very general way, leaving open the possibility that some biologically interesting, but as yet undiscovered quantity increases over time. Second, looking toward historical patterns for definitions of progression is essentially looking at data and then designing an hypothesis to fit. Progression ought to be defined prospectively rather than retrospectively, meaning that it ought to have predictive value; if evolution is progressive, then we ought to be able to define, \textit{a priori}, a quantity that will increase.

The definition of progression used here was purposefully chosen to deal with the above-mentioned difficulties. Thus, as it stands, it necessarily is not linked to any specific biological measurement. By the definition used here, the quantity that might increase over time need not have any obvious biological interpretation outside of the role that it plays in progressive evolution. This level of generality seems desirable if we are asking questions about the existence of such a quantity without necessarily knowing anything specific about what it might be. Such generality does mean, however, that if evolution is progressive in this sense, then the progressive trait might well be some highly complicated characteristic of the population that does not necessarily correspond to any biological attribute of an organism that is \textit{a priori} natural. In this way, some readers might prefer to view the theorem presented here as a definition of progressive evolution rather than as a statement about the limitation of theory. In other words, we might define progressive evolution as an evolutionary process for which we could, in principle, construct a negation-complete evolutionary theory. The theorem then says that this definition is equivalent to there existing some quantity that increases over evolutionary time.

Decidability results, such as those presented here, are often prone to misinterpretation \citep{Franzen:2005}. Therefore it is important to be clear about what the above theorem says as well as what it does not say. First, the theorem does not imply that developing a predictive theory of evolution is impossible. A very large portion of current research in evolutionary biology is directed towards developing such predictive capacity and therefore the theorem takes the existence of such a theory as a starting point. The rationale is to determine whether there might still be other, inherent, limits on the kinds of questions that can be answered even if we are successful in pushing the development of current research in this direction. The theorem demonstrates that there are such inherent limits, and in essence the problem arises from a difficulty in predicting the places that evolution does not go. In other words, although a predictive theory can always be used to map out the course of evolution, interestingly, it cannot always be used to map out the courses that evolution does not take. The theorem presented here, in effect, demonstrates that doing the latter is not possible unless evolution is progressive.  

How are these considerations to be interpreted in the context of examples like that of influenza evolution discussed earlier? First, as already mentioned in that example, the analysis would begin by taking what is essentially a best-case scenario, and supposing that we have enough knowledge of the system to develop an open-ended model that perfectly predicts (possibly in a probabilistic way) the genetic composition of the influenza population in the next time step, as a function of its current composition. Then we ask, is there a significant probability that another flu pandemic with the 1918 strain will ever occur? The above theorem states that, even if we had such a perfect model, this kind of question is unanswerable unless influenza evolution is progressive. In other words, unless some characteristic of the influenza population changes directionally during evolution (e.g., some aspect of the antigenicity profile changes directionally) such a prediction will not be possible. Moreover, this limitation arises because, even though we can use our perfect model to map out the course of influenza evolution over time, this need not be enough to map out the parts of genotype space that influenza will \textit{not} explore.

The above limitations apply to predictions about the genetic evolution of the population, but what if we are interested only in phenotypic predictions? For example, could we predict whether or not an influenza pandemic similar in severity to that of 1918 will ever occur again, regardless of which strain(s) cause the pandemic? Likewise, could we predict whether or not resistance to antiviral medication will ever evolve, regardless of its genetic underpinnings? If the genotype-phenotype map is one-to-one, then predicting phenotypic evolution will be no different than predicting genotypic evolution. Even if many different genotypes can produce the same phenotype, however, predicting phenotypic evolution still involves predicting whether or not certain subsets of genotype space are visited during evolution. As a result, all of the aforementioned limitations should still apply to such cases. The only exception is if the genotype-phenotype map resulted in the dimension of phenotype space being finite even though the dimension of the genotype space was effectively infinite. Even in this case, however, the above limitations to prediction would still apply unless phenotypic knowledge alone was sufficient to predict the state of the population from one time step to the next (i.e., if we didn't need to consider genetic state to understand evolution). While this might be possible for some phenotypes of interest, it seems unlikely that it would be possible for all possible phenotypes.

One might argue, however, that some patterns of phenotypic evolution are very predictable. For example, the application of drug pressure to populations seems inevitably to lead to the evolution of resistance to the drug. How are these sorts of findings reconciled with the results presented here? First, although the evolution of resistance does appear to be somewhat predictable, we must distinguish between inductive versus deductive predictions. One reason we feel confident about predicting the evolution of drug resistance is that we have seen it occur repeatedly. Therefore, by an inductive argument we expect it to occur again. Such inductive predictions are conceptually similar to extrapolating predictions from a statistical model beyond the range of data available.  On the other hand, deductive predictions are made by deducing a prediction from an underlying set of principles or mechanistic processes. In a sense, inductive predictions require no understanding of the phenomenon in question whereas deductive predictions are based on some underlying model of how things work. The results presented here apply solely to deductive predictions.

A second possibility with respect to the evolution of things like drug resistance, however, is that evolution is progressive (at least at this `local' scale). For example, it might well be that if we formulated an accurate underlying model for how influenza evolution proceeds in the presence of antiviral drug pressure, there would be some population-level quantity that changes in a directional way during evolution. Indeed it seems plausible that it is precisely this kind of directionality that makes us somewhat confident we can predict evolution in such cases. It should be noted, however, that even if evolution if not progressive the theorem presented here does not rule out the possibility that some predictions can be made. For example, it is entirely possible that a theory could still be developed to make negation-complete predictions about the evolution of drug resistance. The theorem simply says that it will not be possible to make negation-complete predictions about any arbitrary aspect of evolution unless the evolutionary process is progressive. 

As already mentioned, all of the results presented here begin with the assumption that we can develop a theory to predict evolution from one time step to the next. Whether or not current theoretical approaches can be pushed the point where this is true remains a separate, and open, question. There are certainly considerable obstacles to doing so unless the evolutionary system of interest is very simple (e.g., \cite{Ibarra:2002}). In addition to the problem that historical contingencies raise, the role of uncertainty in initial conditions, much like those in weather forecasting, might preclude long-term predictions (although probabilistic statements might still be possible). This remains an important and active area of research on which the theorem presented here offers no perspective. Rather it simply reveals that, in the event that theory is eventually developed to do so, it will still face inherent limitations on the kinds of questions it can answer unless evolution is progressive. 

Although a negation-complete theory for the entire evolutionary process of interest is not possible unless evolution is progressive, this also does not preclude the possibility that a perfectly acceptable, negation-complete, theory might be developed for short-term and/or local predictions. Indeed, just as similar inherent limitations in computability theory and mathematical logic have not prevented people from making astonishing progress in these areas of research, so to is the case for evolutionary biology. As mentioned in the introduction, many theoretical advances have already been made by focusing on subsets of the space of potential evolutionary outcomes. Continuing to push theoretical development in this direction by broadening the space considered will be possible regardless of the nature of the evolutionary process. The theorem does imply, however, that unless evolution is progressive, it will not be possible to encompass all such developments within a single unified set of principles from which all negation-complete evolutionary predictions can be drawn.

There are some previous theoretical results in the literature that consider the extent to which evolution exhibits a directional tendency and it is useful to consider how the present results relate to these previous works. For example, it has been shown previously with quite general stochastic models of evolution that a quantity termed `free fitness' is always non-decreasing during evolutionary change \citep{Iwasa:1988}. The analysis, however, did not allow for open-ended evolution because the state space was assumed to be finite, and the Markov model used was (implicitly) assumed to be positively recurrent. As a result, a unique stationary distribution existed and thus continual evolution was precluded. 


It might be reasonably argued however that, although analyses such as \citep{Iwasa:1988} do not allow for truly open-ended evolution, if the state space is large enough, and if the transient dynamics are long enough, then it is effectively an open-ended model. As such, should not the results with respect to free fitness still apply? In other words, does this not then suggest that there is some quantity (free fitness) that increases during evolution, and thus that a negation-complete theory is possible? The answer is no, and the reason is subtle but important. The definition of free fitness in \citep{Iwasa:1988}, like other quantities that have been suggested to change directionally during evolution (e.g., \cite{Adami:2000}) are based on measures closely related to entropy. Importantly, the mapping between these measures of entropy and population states is not one-to-one because there are many (indeed, potentially infinitely many) biologically distinct population states that have the same value of entropy (or the same value of `free fitness'). As a result, even though measures like free fitness might not decrease during evolution, an indefinite amount of biologically interesting and significant evolutionary change can still occur without any change in free fitness. Roughly speaking, although measures related to things like entropy provide an interesting physical quantity that might change directionally, the relationship between entropy and quantities that are of biological interest need not be simple.

In a similar vein one might argue that, because biological evolution takes place within a physical system that is subject to the Second Law of Thermodynamics, ultimately a general measure entropy must provide a directionality to the system. Again, while this is true is terms of the system as a whole, the mapping between entropy and the population states of biological interest is not one-to-one. Thus, even though the total entropy of the entire physical system must always increase, the entropy of any component part (e.g., the biological part of interest) need not change in this way.

What do all these considerations have to say about how the process of evolution is studied, or how current theoretical research is done? Should evolutionary biologists care about such results? For instance, do the results point to new ideas that might help us do theory better? Although there is no single answer to this question, there are two points worth making in this regard. First, the distinction between open and closed-models seems like a useful, and currently somewhat under-appreciated, way to categorize models of evolution. As such it does suggest some new directions in which evolutionary theory might be taken, particularly given that open-ended models are sometimes amenable to asking novel, and potentially very important, evolutionary questions that cannot be addressed with closed models (e.g., \cite{Fontana:1994, Fontana:1998b, Lenski:1999, Stadler:2001, Yedid:2001, Wilke:2001, Yedid:2002, Lenski:2003, Chow:2004, Ostrowski:2007}). Second, to the extent that one cares about developing theory for open-ended evolutionary processes, the theorem presented here then reveals that there is an inherent `upper bound' on how far we can push the predictive capability of such theory. In particular, although such theory opens the door to asking new evolutionary questions, unless evolution is progressive, there will remain some such questions that are unanswerable. Furthermore, although it will likely be difficult to use the theorem as a means of proving that evolution is progressive (i.e., by developing a negation-complete theory) or to use the theorem to prove that a complete evolutionary theory is possible (i.e., by determining that evolution is progressive) the result does nevertheless reveal that these two important, and somewhat distinct, biological ideas are fundamentally one and the same thing.

My intention was not to imply that the theorem could be used to determine decidability from knowledge of progression, or the reverse. Rather, it was to prove (within the set of assumptions used) that decidability and progression can be viewed as one of the same thing. 

The theorem presented here has close ties to G\"odel's Incompleteness Theorem for axiomatic theories of the natural numbers \citep{Godel:1931, Nagel:1958, Davis:1965, Heijenoort:1967, Smith:2007}. An axiomatic theory consists of a set of symbols, a logical apparatus (e.g., the predicate calculus), a set of axioms involving the symbols, and a set of rules of deduction through which new statements involving the symbols can be derived (termed `theorems'; \cite{Smith:2007}). Given such a system, theorems can be derived through the repeated algorithmic application of the rules of deduction. 

In the early 1900's there was a concerted attempt to produce such an axiomatic theory that was meant to represent the natural numbers, with the proviso that it yield all true statements about the natural numbers, and no false ones; \citep{Whitehead:1910, Smith:2007}. G\"odel's Incompleteness Theorem \citep{Godel:1931, Nagel:1958, Davis:1965, Heijenoort:1967, Smith:2007}, however, revealed that this is impossible for any axiomatic system sufficiently rich that it can make simple number-theoretic statements. For example, it shows that if the axiomatic system is rich enough that it can express the number-thoeretic statement corresponding to the predicate `$x \in R_E$', then it cannot produce all true number-theoretic statements and no false ones \citep{Smith:2007}. For if it could, then it could always produce the number theoretic statement corresponding to either `$x \in R_E$' or `$x \notin R_E$' as a theorem, because one of the two must be true. But if it can do this, then it provides an algorithmic procedure for deciding the predicate `$x \in R_E$', and we know that this is not always possible as the results presented here illustrate. 

The Halting Problem from computability theory \citep{Turing:1936, Cutland:1980} is also intimately related to the results presented here. As already detailed, the question of whether a population state is evolutionarily attainable is equivalent to the question of whether a given positive integer is in the range of a particular computable function. Moreover, this latter question is directly connected to the analogous question of whether a given integer is in the domain of a computable function (i.e., whether, given a particular integer input, the function returns a value in finite time). The latter problem is precisely the Halting Problem, and it is known that there is no general algorithmic procedure for solving the Halting problem for arbitrary computable functions \citep{Turing:1936, Cutland:1980}. 

As mentioned earlier, in a very general sense, the results presented here are applicable to any system that can be faithfully described by a Markov dynamical system over an infinite set of discrete possibilities (i.e., an open-ended dynamical system). Therefore, one might ask whether there is anything in the results presented that is particular to evolution \textit{per se}? In one sense the answer is `no', but therein lies the power of such mathematical abstraction; it reveals the underlying, key, structure of the process. Evolution will be an open-ended dynamical system whenever heredity is indefinite, and it therefore shares a fundamental similarity with all other processes that are also such open-ended dynamical systems.

At the same time, however, the results do have special significance for evolution. There are, perhaps, relatively few other kinds of processes of interest that share the property of being such an open-ended dynamical system in a meaningful way. For example, a great many processes of interest have a relatively small space of potential outcomes, and are thus clearly not open-ended. Furthermore, for those processes that are potentially open-ended, it is sometimes of little theoretical interest to distinguish among all possible outcomes, and therefore the space of relevant outcomes can still be relatively small. Moreover, even when the space of potential outcomes of interest truly is open-ended, some processes (e.g., some physical processes) obey simple enough dynamics that such negation-complete predictions can readily be made (i.e., the system is `progressive' is the sense considered here). Thus, the limitations detailed by the theorem are of interest, primarily for those processes that are both open-ended, and that are complex enough that the question of progression is unresolved (Appendix \ref{re}). Evolution under indefinite heredity might be a somewhat unique process in satisfying both of these criteria. 

There are, however, other processes of interest for which such decidability results might be of interest. After all, in an important sense, biological evolution is nothing more than the emergent properties of physics and chemistry. In fact such limitations on theory have been discussed previously, particularly as they relate to the so-called theory of everything in physics \citep{Hawking:website}. It is probably safe to say that no general concensus on this issue has yet been reached \citep{Franzen:2005}; however, the theorem presented here has implications for any physical or chemical theory that aims to explain evolutionary phenomena. It demonstrates that a rational, deductive, approach to such theory will necessarily face some inherent limitations on the answers that it can provide.


\section*{Acknowledgements}

I thank N.H. Barton, G. Bell, S.A. Frank, S. Gandon, A. Gardner, S. Gavrilets, B. Glymour, M. Hochberg, L. Nagel, S.P. Otto, S. Proulx, A. Read, A. Rosales, P. Taylor, M. Turelli, and the Queen's Math Bio Group for inspiring conversations and feedback on the manuscript that greatly helped to clarify the results and to put them into perspective. This research was supported by an NSERC Canada Discovery Grant, an E.W.R. Steacie Fellowship, and the Canada Research Chairs Program. 

\newpage






\bibliographystyle{amnat}
\bibliography{Godel}

\newpage

Figure 1: A schematic representation of the coding of population states, and the theorem. Middle irregular shape represents the space of population states, $S$, with four states depicted (the ovals). Roman numerals indicate the time when each state is visited during evolution (silver-shaded state, $s=\{T,T,T\}$, is never visited). Vertical ovals on right and left represent two different codings by the positive integers, along with their respective evolutionary mappings, $\phi_E(n)$ and $\phi_{\hat{E}}(n)$, over the first three time steps. If evolution is progressive, then Coding 2 is possible, and the theorem then says we can `decide' any population state, $s \in S$. For example, we can decide state `T,T,T' by finding its code (i.e., `1'), and then iterating the map, $\phi_{\hat{E}}(n)$, until we obtain an output greater than `1' (this occurs at time step 1 because $\phi_{\hat{E}}(1)=2$). If `1' has not yet been visited by this time, it never will be. Conversely, if all population states are decidable, then under Coding 1 we can apply the algorithm provided in Part 2 of the theorem's proof to obtain Coding 2, thereby demonstrating that evolution is progressive.

\newpage
\section*{Appendices}

\section{Theory} \label{theory}

The term `theory' is used in a technical sense. A theory consists of a set of symbols that constitute the language of the theory, a set of premises which are taken as given, and a set of rules of inference \citep{Smith:2007}. The symbols represent certain components of reality, and the premises constitute statements about reality through the interpretation of the symbols. The rules of inference then constitute valid ways of deducing new statements about the symbols of the language, and thus through interpretation, new statements about reality. Thus, within such a theory, statements are derived by taking some premise(s), and applying the rules of inference. 

Statements derived through a series of deductive arguments using the rules of inference are referred to as theorems of the theory. The result of the main text is valid for any evolutionary theory whose theorems are recursively enumerable (Appendix \ref{comp}); i.e., any theory whose theorems can be derived through the use of a finite (but possible large) number of mechanical, or algorithmic, steps (e.g., as laid out in the rules of inference; Appendix \ref{comp}). This is clearly true for any such theory based on computation, since computers do nothing more than mechanically follow rules \citep{Cutland:1980}. It is also true for any axiomatic theory, since the theorems of any such theory can be derived simply by applying the mechanical rules of inference to the axioms \citep{Smith:2007}.

A great deal of current quantitative theory in evolutionary biology fits the above template. For example, current theory often abstracts reality mathematically by assigning formal symbols to things like allele frequencies and population sizes. A set of premises is then taken, for example, by formalizing an hypothesis about how genotypic fitnesses are determined. Next, a finite number of applications of `rules of inference' are used (e.g., the application of certain mathematical operations) in order to derive statements about the formal symbols of this theory. Finally, these symbolic statements are then interpreted again in terms of their biological meaning, and hence predictions about evolution are made (Fig. S1).

\textit{Figure S1: A schematic representation of the relationship between the biological process of evolution and theory. The example given illustrates classical population-genetic theory. A formal system is created to represent elements of evolution (e.g., $p(t)$ represents the number of the blue genotype at time $t$). A set of premises is specified (e.g., initial genotype numbers, how genotypic fitnesses are determined, etc. - this is embodied by the mapping $F$). Rules of deduction are then followed (e.g., repeated application of the mapping $F$) to obtain new statements about elements of the formal theory (e.g., $p(1), p(2), p(3)$ etc.). These new elements are then interpreted in terms of evolution (e.g., as predictions about genotype numbers at future times).}

\section{Some results from computability theory} \label{comp}

A function is computable if it can be evaluated by an Unlimited Register Machine (URM) in a finite numbers of steps \citep{Cutland:1980}. The Church-Turing Thesis states that any function we might view as being evaluated through a mechanical procedure can be evaluated by a URM \citep{Cutland:1980}. Thus, given the Church-Turing Thesis, the easiest way to ascertain whether something is computable is to consider whether a computer could be programed to do it in such a way that an output is guaranteed, in a finite (but possibly very large) number of steps.

\textbf{Definition}: A function is \underline{total} if it is computable over all natural numbers.

\textbf{Definition}: A function is \underline{partial} if it is computable only over some (nonempty) subset of the natural numbers.

\textbf{Definition}: A set is \underline{denumerable} if there exists a bijection between it
and the natural numbers.

\textbf{Definition}:  A set is \underline{effectively denumerable} if this bijection, and its
inverse, are computable.

\textbf{Definition}: The \underline{characteristic function} of a set of natural numbers, $A$, is

\begin{eqnarray}
c_A(n)= \left\{ \begin{array}{cc}
1 & \mbox{if $n \in A$}\\
0 & \mbox{if $n \notin A$}\\
\end{array}\right.
\end{eqnarray}

\textbf{Definition}:  The predicate `$n \in A$' is \underline{decidable} if its characteristic function is computable.

\textbf{Definition}:  The set $A$ is \underline{recursive} if the predicate `$n \in A$' is decidable.

\textbf{Definition}: The \underline{partial characteristic function} of a set of natural numbers, $A$, is

\begin{eqnarray}
\bar{c}_A(n)= \left\{ \begin{array}{cc}
1 & \mbox{if $n \in A$}\\
\mbox{undefined} & \mbox{if $n \notin A$}\\
\end{array}\right.
\end{eqnarray}

\textbf{Definition}:  The predicate `$n \in A$' is \underline{partially decidable} if its partial characteristic function is computable for $n \in A$.

\textbf{Definition}:  The set $A$ is \underline{recursively enumerable} (denoted r.e.) if the predicate `$n \in A$' is partially decidable.

Note that every recursive set is r.e. but not vice versa. Furthermore, a set $A$ is recursive if, and only if, both $A$ and its complement $A^c$ are r.e.. Finally, note that any finite set of numbers is recursive \citep{Cutland:1980}.

The following concepts and notation will also prove useful: 

First, because any computable function can be evaluated through a series of steps, we can define $c^o_A(n)$ as the value of $c_A(n)$ after the $o^{th}$ step in its evaluation. In particular, $c^o_A(n)$ evaluates to `null' if it has not returned a value by the $o^{th}$ step. 

Second, a standard result from computability theory demonstrates that there exists a computable bijection between $\mathbb{N}^+$ and $\mathbb{N}^+ \times \mathbb{N}^+$ \citep{Cutland:1980}. We will denote this mapping by $B:n \mapsto (T_1(n),T_2(n))$.

Third, the notion of an `unbounded search' is central in computability theory. In particular, it is standard to use the notation $\mu y(f(y)=k)$  to denote `the smallest value of $y$ such that $f(y)=k$'. 

Fourth, a fundamental theorem of computability theory demonstrates that the set of all computable functions is denumerable \citep{Cutland:1980}. Thus, we can use $\phi_k(n)$ to denote the $k^{th}$ computable function, and $R_k$ and $D_k$ as its range and domain respectively. We will also make use of the notation $R_k(n)=\{x:\phi_k(i)=x, i \leq n \}$. In other words, if $\phi_k(n)$ is evaluated for increasing values of $n$, then $R_k(n)$ is the subset of the range of $\phi_k(n)$ that has been visited by step $n$. This is clearly computable for any $n$ if $\phi_k(n)$ is total.

Finally, notice that it was implicitly assumed that the mapping, $G$ corresponding to the evolutionary process is computable, and thus $E(n)$ is a computable function. Thus, the evolutionary process is, in an important way, nothing other than computation. Although it is not practically feasible to verify or refute this assumption for most evolutionary systems, there are very good reasons to expect that this assumption is reasonable. First, if we are willing to view the processes occurring in our biological system as being purely `mechanical', then we can appeal to the Church-Turing Thesis to argue that $G$ must thereby be computable. Second, the use of the term `evolution', as a process, should not be restricted to a particular instantiation of this process, as for example occurs in carbon-based life. For example, there are very good reasons to think that the processes occurring in \textit{in silico} evolution are fundamentally the same as those occurring in biological evolution. As such these would clearly be computable. Finally, even if biological evolution isn't formally computable (i.e., it is not ÔmechanicalÕ) we nevertheless usually proceed by assuming that it can be modeled using computation.

\section{The set of population states is effectively denumerable} \label{popstate}

Here we prove that the set of possible population states is effectively denumerable; i.e., that there exists a computable bijection between the population states and the positive integers with a computable inverse. Such sets are also called effectively denumerable.

Proof: We simply need to demonstrate an effective procedure (i.e., a computable procedure) for both encoding and decoding the population states into positive integers. Let $M$ be the maximum possible population size (a positive integer). Each of the M `slots' is either vacant, or filled by an individual that is completely characterized by its DNA sequence. Furthermore, we can set A=0, C=1, G=2, T=3, and then read the DNA sequence from its 5' to 3' end, thereby establishing a unique characterization of each slot in the population. 

(A) Encoding: For each of the $M$ slots calculate a numeric code as follows: Reading the DNA from its 5' to 3' end, for the $n^{th}$ base, take the $n^{th}$ prime number and raise it to the power corresponding to this base as listed above. Multiply all these numbers together. This gives a unique number for each distinct DNA sequence, and thus the mapping is injective. Furthermore, since all positive integers greater than or equal to 2 have a unique prime factorization, all such integers correspond to a DNA sequence. Thus, if we code the state `vacant' with the number 1, the mapping is surjective as well. Furthermore, this procedure is computable for any piece of DNA. This shows that there is a computable encoding for each slot, and since the population is simply the union of a finite number of such slots, the population state has a computable encoding as well. In particular, the coding of each slot locates a point in $\mathbb{N}^+ \times \cdots \times \mathbb{N}^+$ (where $\mathbb{N}^+$ appears $M$ times) that can be uniquely identified by its indices. One can then cycle through all possible indices as follows: start with all indices that sum to 1, then those that sum to 2 etc. This is computable, and for each instance we simply assign a code number in increasing order. 

(B) Decoding: For any given code number, cycle through the sets of indices as above, stopping once the code number is reached, and determine those indices. Once these indices have been obtained, one can determine their corresponding DNA through their prime factorization.



\section{Some additional technical information about the theorem} \label{re}

The theorem of the text would be of little interest if it were never possible for `$x \in R_E$' to be undecidable. It is well-known in computability theory that there exist computable functions for which such predicates are undecidable (\citep{Cutland:1980}; Appendix \ref{re}), but the evolutionary process considered represents a special kind of computable function. In particular, it must satisfy the mapping $\phi_k(n+1)=G \big (\phi_k(n) \big )$ for all $n$, where $G()$ is a computable function with appropriate domain. The subset of computable functions satisfying this relation will be referred to as Markov, total, computable functions. 

This section presents a series of three lemmas that, together, demonstrate that there do in fact exist Markov computable functions for which `$x \in R_E$' is undecidable (see also \cite{Cutland:1980, Smith:2007}). In such cases, the set of evolutionarily attainable states, $R_E$ will be called `recursively enumerable' (r.e.; because `$x \in R_E$' is always at least partially decidable for Markov computable functions). On the other hand, if  `$x \in R_E$' is decidable, then $R_E$ is said to be `recursive' (Appendix \ref{comp} and Appendix \ref{re}). 

\emph{Lemma 1: A set of numbers is recursively enumerable if, and only if, it is the range of some total, computable, function. Note: we could relax the `total' requirement without much change.}

Proof:
(i) $A \mbox{ r.e.}  \Rightarrow$ `$A$ is the range of a total computable function'

Given $A$ is r.e., the partial characteristic function of $A$ is computable; i.e., 

\begin{eqnarray}
\bar{c}_A(n)= \left\{ \begin{array}{cc}
1 & \mbox{if $n \in A$}\\
\mbox{undefined} & \mbox{if $n \notin A$}\\
\end{array}\right.
\end{eqnarray}

is computable. 
Now first choose an $a \in A$. This is a computable operation since we can simply use the bijection $B:n \mapsto (T_1(n),T_2(n))$ to evaluate $\bar{c}^{T_2(n)}_A(T_1(n))$ for increasing $n$ until it returns a value of 1, and then identify the corresponding value $T_1(n)$. Next, we can define the computable function

\begin{eqnarray}
g(x,o)= \left\{ \begin{array}{cc}
x & \mbox{if $\bar{c}^o_A(x)=1$} \\
a & \mbox{otherwise}\\
\end{array}\right.
\end{eqnarray}

Then, again we can use the computable bijection $B:n \mapsto (T_1(n),T_2(n))$ to define $f(n) = g \big(T_1(n),T_2(n)  \big)$. This is a total computable function with range equal to $A$. 

(ii) `$R_k$ is the range of a total computable function' $ \Rightarrow R_k \mbox{ r.e.}$ 

Consider the total function $\phi_k(n)$. We can then construct the computable partial characteristic function for $R_k$ as follows: For any input value, $x$, output the value 1 after evaluating $\mu i (\phi_k(i)=x)$. 

Q.E.D.

Given Lemma 1, we can then prove the following, second, lemma;

\emph{Lemma 2: There exists total computable functions whose ranges are r.e. but not recursive.}

Using Lemma 1, we can prove Lemma 2 by proving that there exist sets that are r.e. but whose complements are not r.e.

\emph{Proof Sketch (by construction) see \cite{Smith:2007}: }

We will demonstrate that $K=\{n: n \in R_n \}$ is one such set. It is clear, therefore, that other such sets can be constructed as well.

First it can be proven that $K^c$ is not r.e. using Cantor's diagonal argument (e.g., see \cite{Smith:2007}). In particular, since all r.e. sets are the range of some computable function, and since the computable functions are denumerable, the set of all r.e. sets is denumerable. So we simply need to construct a set that is not in this list. Choosing numbers $n$ such that $n \notin R_n$ satisfies this property, and this is exactly $K^c$. 

All that remains then is to show that $K$ is r.e. As with characteristic functions, all computable functions are evaluated through a series of operations for each input, and therefore we can consider the $o^{th}$ operation of any computable function. Therefore, define

\begin{eqnarray}
g(x,o,n)= \left\{ \begin{array}{cc}
\phi_n(x) & \mbox{if $\phi_n(x)$ \mbox{ halted by operation $o$ in its evaluation}}\\
n+1 & \mbox{otherwise}\\
\end{array}\right.
\end{eqnarray}

This is a computable function. Now we can use the bijection $B:n \mapsto (T_1(n),T_2(n))$ to define $f(z,n) = g \big(T_1(z),T_2(z),n  \big)$. This is also computable, and for any given $n$ and $z$ it outputs either $n+1$ or else an element of $R_n$. We can then construct the computable partial characteristic function for $K$ as follows: For any input value, $n$, output the value 1 after evaluating $\mu z (f(z,n)=n)$.

These results show that there exist computable functions whose ranges are r.e. but not recursive. Note that some such functions might have the same output values for more than one value in their domain, but these cannot be Markov computable functions. The reason is simply that the mapping $G$ ensures that, if $R_E$ is infinite, then $\phi_E(n)$ can never repeat itself as $n$ increases (see Lemma 1, Appendix \ref{stoch}). Therefore, we still need to demonstrates that, even if we restrict attention to Markov computable functions, some such functions have r.e. ranges that are not recursive. This is done in the third lemma:

\emph{Lemma 3: For every total computable function having a range that is r.e. but not recursive, there exists a total computable Markov function with the same range.}

Proof: Suppose that $\phi_k(n)$ is total and has an r.e. range that is not recursive (and thus $R_k$ is infinite). Define the computable function $\phi_{\hat{k}}(n)=\phi_k(z(n))$, where $z(n)=\mu i (\phi_k(i) \notin R_k(n-1))$. It is clear that $\phi_{\hat{k}}(n)$ is a total, computable function with range $R_k$. Now we simply need to show that $\phi_{\hat{k}}(n+1)=G \big ( \phi_{\hat{k}}(n)  \big )$ for all $n$ for some computable $G()$. By construction we can see that the computable function $G(y) = \phi_{\hat{k}}( \mu z (\phi_{\hat{k}}(z)=y)+1   )$ works, where its domain is $R_k$. This function takes a state $y$, finds the unique time at which this state occurs (i.e., $\mu z (\phi_{\hat{k}}(z)=y)$ - this is computable), and then adds 1. The resulting value is then used in the function $\phi_{\hat{k}}(n)$ to compute the state in the next time step. In particular, we can see that $G \big ( \phi_{\hat{k}}(n)  \big )=\phi_{\hat{k}}(n+1)$.

Q.E.D.

\section{Continuous Time \& Stochasticity} \label{stoch}

For simplicity of exposition, all results of the main text have assumed that the evolutionary process is deterministic and that generations are discrete. Here we show that an analogous theorem holds if we relax these restrictions. 

To begin, it is easy to see that the assumption of discrete generations is immaterial. In particular, if we take generations to be continuous, then we can suppose that, at any instant in time, only a single event is possible (e.g., individual birth or death). Thus, because the state space is discrete, we can simply view the continuous-time process as one in which discrete events occur at points in time that need not be uniformly spaced.

Allowing for stochasticity requires more work. If the evolutionary process is deterministic, then there is a single population state possible for each point in time, $n$. In the analysis of this case, we supposed that we had complete knowledge, not only of the evolutionary mapping, $G$ an its initial condition, but of the solution to this mapping, $\phi_E(n)$ as well (and it is a total, computable, function).  


Now there will be uncertainty in what the population state will be at time $n$, and in fact there will potentially be several different states that the population might attain at $n$. Some of these might be more likely than others in that, if we replayed the evolutionary process multiple times, certain states might arise more often than others. Thus we might imagine a probability distribution over the set of positive integers at each time step, $n$. By analogy with the deterministic case, we make a Markov assumption, meaning that the probability distribution on the population states at any given time, $n$, depends only on the population state in the previous time, $n-1$. In other words, there is some mapping, $H$, from current population state to the probability distribution over the population states in the next time period.  The solution of this mapping (given an initial condition) then gives the probability distribution over the states at each point in time.

Just as with the deterministic case, we suppose that we have complete knowledge of the solution of this evolutionary process in the following sense: at any time $n$, we have a total, computable function that tells us simply the set of states, at that time, that have positive support. Thus,  we have a total, computable, set-valued function $\tilde{\phi}_E(n)$ that gives the set of ``feasible" states at time $n$. The `tilde' signals that this function is now a set-valued function, rather than an integer-valued one. And again the goal of a negation-complete theory would then be to decide whether any given state lies within the set of feasible states or not. 

One objection to this formulation is that we might expect all states have some nonzero probability, even if it is vanishingly small. As such, under this definition all states would then be trivially feasible. There are at least two potential responses to this objection. First, while it is true that many models of evolution assume that all states have nonzero probability (e.g., many stochastic models of mutation-selection balance, including those with an infinite number of different alleles; \cite{Kimura:1964}), this is usually because they are `closed' models in the sense described earlier. In particular they often assume, for mathematical convenience, that the stochastic process is irreducible and positively recurrent. This then implies that a unique stationary distribution exists \citep{Grimmett:1992} and thereby rules out the possibility of open-ended evolution. Although it is possible to develop a model for open-ended evolution that still has nonzero probability for all states, it is not obvious that this need be true of real open-ended evolution. For example, out of the effectively infinite number of different nucleotide combinations that could make up a genotype, we might expect at least some of these to be truly lethal. On a more practical level, given the analysis presented here it seems reasonable to expect that a similar theorem could be proved if we instead defined a state as being feasible if it occured with some probability greater than a small threshold value, $\epsilon > 0$. At this point, however, such a theorem remains conjecture. 

Given that all of our considerations with respect to computability have been restricted to integer-valued functions, we now need to make the notion of computability of $\tilde{\phi}_E(n)$ more precise. The set-valued function $\tilde{\phi}_E(n)$ can be thought of as consisting of two separate computable functions, each of which is an integer-valued function and so fits within the notions of computability already discussed. The first function is simply a computable function $\phi_E(i)$ as before, whose range is now thought of as the set of feasible population states. The argument $i$ here is now no longer meant to be evolutionary time, however, but rather is simply an index whose meaning is described below. The second computable function we denote by $\phi_{E^*}(n)$, and it specifies the number of feasible population states in generation $n$ in the following way: the set of all feasible population states at time 1; i.e., $\tilde{\phi}_E(1)$ is given by $\{\phi_E(1), \phi_E(2), ...,\phi_E(k_1)\}$, where $\phi_{E^*}(1)=k_1$. Likewise, $\tilde{\phi}_E(2)=\{\phi_E(k_1+1), ...,\phi_E(k_1+k_2)\}$, where $\phi_{E^*}(2)=k_2$, and so on. In this way, we can apply the same notions of computability to the set-valued function $\tilde{\phi}_E(n)$ by applying them to its component, integer-valued, functions $\phi_E(i)$ and $\phi_{E^*}(n)$. We will assume that the set $\tilde{\phi}_E(n)$ is finite for all $n$, which guarantees that it be computable. Nevertheless, it seems reasonable to expect that some formulations in which this set is infinite would still be computable, and thus would still fit within the results that follow. 

As in the deterministic case, we must also specify the initial conditions, in addition to the mapping, $H$. Then, in terms of the mapping, $H$, if $x \in \tilde{\phi}_E(n)$ is a feasible population state at time $n$, the set of feasible population states at time $n+1$ is given by $\tilde{\phi}_E(n+1) = \bigcup_{x \in \tilde{\phi}_E(n)} \text{support} H(x)$, where $\text{support} H(x)$ denotes the set of states for which $H(x)$ has positive support. The range of $\tilde{\phi}_E(n)$ is the set of all states that are feasible at some time (i.e., it is the range of $\phi_E(i)$). Likewise, a state is evolutionarily attainable if there is some time for which it is feasible. A complete evolutionary theory is one for which the predicate `$x \in R_E$' is decidable; i.e., if, given any population state, we can decide whether it is feasible at some time.

The same definition of progressive evolution can be used in both the deterministic and stochastic cases. To specify this precisely, we need the following Lemmas;

\emph{Lemma 1: In the deterministic case, a new state is visited every time step if, and only if, evolution is unbounded (i.e., $R_E$ is infinite)}

\emph{Lemma 2: In the stochastic case, at least one new state is feasible every time step if, and only if, evolution is unbounded (i.e., $R_E$ is infinite)}

Proof is given of Lemma 2 only (Lemma 1 can be proven in an analogous fashion). We note that, in the remainder of this section, we use the notation $R_E(n)$ to denote the set of population states that have been visited (i.e., feasible) by step $n$ of the set-valued function, $\tilde{\phi}_E(n)$ (i.e., not step $n$ of $\phi_E(n)$). Equivalently, it denotes the range of $\phi_E(i)$ visited by step $i=k_1+k_2+ \cdots + k_n$.

Proof:

`At least one new state is feasible each time step' $\Rightarrow$ `Evolution unbounded'

This direction of the implication is obvious since, if at least one new state is feasible each time step, then the fact that  $\tilde{\phi}_E(n)$ is total implies that $R_E$ is infinite.

`Evolution unbounded' $\Rightarrow$ `At least one new state is feasible each time step'

Contrary to the assertion, suppose instead that $R_E$ is infinite but that there is some time, $n^*$ at which no new state is feasible. In other words, for some time $n^*$, the set $\tilde{\phi}_E(n^*)$ satisfies $\tilde{\phi}_E(n^*) \subseteq R_E(n^*-1)$. The set of feasible states in the next time step is then given by $\tilde{\phi}_E(n^*+1) = \bigcup_{x \in \tilde{\phi}_E(n^*)} \text{support} H(x)$. Furthermore, for each element, $x \in \tilde{\phi}_E(n^*)$, $\exists n_x<n^*$ such that $x \in \tilde{\phi}_E(n_x)$  (from the hypothesis that $\tilde{\phi}_E(n^*) \subseteq R_E(n^*-1)$). Therefore, for each element, $x \in \tilde{\phi}_E(n^*)$, we have that $\text{support} H(x) \subseteq \tilde{\phi}_E(n_x+1)$, where $n_x < n^*$. Thus, we have 

\begin{eqnarray}
\tilde{\phi}_E(n^*+1)  &=&  \bigcup_{x \in \tilde{\phi}_E(n^*)} \text{support} H(x) \\
  & \subseteq & \bigcup_{x \in \tilde{\phi}_E(n^*)} \tilde{\phi}_E(n_x+1)\\
  & \subseteq & R_E(n^*-1).
\end{eqnarray}

Hence, by induction, $R_E \equiv R_E(n^*-1)$, which is finite, yielding a contradiction.

Q.E.D.

Notice that, in the deterministic case, when evolution is unbounded the computable function $\phi_E(i)$ never repeats a previously attained value as $i$ increases (Lemma 1 above). In the stochastic case, however, even when evolution is unbounded, $\phi_E(i)$ \textit{can} repeat previously attained values as $i$ increases. The key connection between the two cases is that, in the stochastic case, $\phi_{E^*}(n)$ is such that, when the outputs of $\phi_E(i)$ are grouped into their corresponding evolutionary generations, each such grouping always contains at least 1 new feasible state (Lemma 2 above).

Now, returning to the proof of the theorem, in the deterministic case, Lemma 1 shows that a new population state is visited at every time step. And if evolution is progressive, then there is some way to recode the populations states such that, the code number of these new states that are visited over time increases. Likewise, Lemma 2 shows that at least one new population state becomes feasible at every time step, although some visited population states might have been visited previously as well. Nevertheless, we still say that evolution is progressive if there is some way to recode the populations states such that, the code number(s) of the new states that become feasible each time step, increases with time. Formally, if we define $\sigma_{\hat{E}}(n) = R_{\hat{E}}(n) \setminus R_{\hat{E}}(n-1)$ as the set of newly feasible states in generation $n$, and $\min \sigma_{\hat{E}}(n)$ as the smallest of these, then evolution is progressive if there exists a computable bijection, $\hat{C}$, between the positive integers and the population states, such that $\min \sigma_{\hat{E}}(n+1) > \min \sigma_{\hat{E}}(n)$ for all $n$. Since the set $R_{\hat{E}}(n)$ is finite and computable for all $n$, $\min \sigma_{\hat{E}}(n)$ is a total computable function.



The proof of the theorem then goes through as follows:

\emph{Theorem: $x \in R_E$' is decidable (i.e., $R_E$ is recursive) if, and only if, there exists a computable one-to-one coding of the population states by positive integers, $\hat{C}$, such that, for the corresponding description of the evolutionary process, $\tilde{\phi}_{\hat{E}}(n)$, 
$\min \sigma_{\hat{E}}(n+1) > \min \sigma_{\hat{E}}(n)$ for all $n$.}

Proof:

Part 1: $\exists \hat{C} \mbox{ s.t.} \min \sigma_{\hat{E}}(n+1) > \min \sigma_{\hat{E}}(n) \mbox{ } \forall n$ $\Rightarrow$ $R_E$ recursive

By hypothesis there exists a computable bijection $\hat{C}$ such that $\min \sigma_{\hat{E}}(n+1) > \min \sigma_{\hat{E}}(n)$ for all $n$. Now for any population state, $x$, in the original coding, let $\hat{x}$ be the corresponding code under bijection $\hat{C}$. Define $z(\hat{x})=\mu i (\min \sigma_{\hat{E}}(i)  \geq \hat{x})$. Clearly `$\hat{x} \in R_{\hat{E}}(z(\hat{x}))$' is decidable since $ R_{\hat{E}}(z(\hat{x}))$ is finite and enumerable. Furthermore $\hat{x} \in R_{\hat{E}}(z(\hat{x})) \Leftrightarrow \hat{x} \in R_{\hat{E}}$ owing to the progressive nature of evolution. Therefore, `$\hat{x} \in R_{\hat{E}}$' is decidable as well. Finally, using $S$ denote the set of population states that are evolutionarily attainable, we have that $\hat{x} \in R_{\hat{E}} \Leftrightarrow \hat{C}^{-1} \hat{x} \in S \Leftrightarrow C\hat{C}^{-1} \hat{x} \in R_E$. Noting that, by definition, $x=C\hat{C}^{-1} \hat{x}$, we obtain $\hat{x} \in R_{\hat{E}} \Leftrightarrow x \in R_E$. Thus, `$ x \in R_E$' is decidable as well.

Part 2: $R_E$ recursive $\Rightarrow$ $\exists \hat{C} \mbox{ s.t.} \min \sigma_{\hat{E}}(n+1) > \min \sigma_{\hat{E}}(n) \mbox{ } \forall n$

We can construct the required computable bijection to show that evolution is progressive as follows.

Since $R_E$ is recursive, we know that `$x \in R_E$' is decidable. So take the population states, $x$, in order and go down the list using the following algorithm:

(i) if $x \notin R_E$ and it is the $k^{th}$ such state up to that point, return the $k^{th}$ odd number.

(ii) if $x \in R_E$, and if it has not yet been assigned a new code number, do the following:

\begin{itemize}

\item calculate $\mu i (x \in \tilde{\phi}_E(i))$ (i.e., the first time that $x$ becomes feasible).

\item calculate $\sigma_E(i)$, the entire set of newly feasible states at $i$.

\item using the notation $|A|$ to denote the cardinality of $A$, assign codes to all of the $|\sigma_E(i)|$ elements in $\sigma_E(i)$, by starting with the $|R_E(i-1)|+1$ even number, up to the $|R_E(i)|$ even number, in any order.

\item move on to the next state in the list.

\end{itemize}

Thus, $R_{\hat{E}}$ is again the set of even numbers, and the new states that are feasible each time step always have larger code values as time increases. In particular, using $\hat{C} C^{-1}$ to denote the algorithm described above in points (i) and (ii), where $C^{-1}$ is the inverse mapping of the coding that generated $x$ (i.e., it takes code $x$ and returns the corresponding population state, $s$), we have $\min \sigma_{\hat{E}}(n+1) = \min \hat{C} C^{-1} \sigma_E(n+1) = 2 |R_{E}(n)+1|$. The last equality follows from the fact that $\hat{C} C^{-1}\sigma_E(n+1)$ determines the first time that each element of $\sigma_E(n+1)$ occurs (which is $n+1$ for all such elements by definition), and then assigns the codes $2 |R_E(n)+1|$ up to $2 |R_E(n+1)|$ for these elements. The minimum of these codes is, of course, $ 2 |R_{E}(n)+1|$ giving $\min \sigma_{\hat{E}}(n+1) = 2 |R_{E}(n)+1|$. As a result, $\min \sigma_{\hat{E}}(n+1)> \min \sigma_{\hat{E}}(n)$ because $|R_{E}(n)|$ is strictly increasing with $n$ (from Lemma 2).

Q.E.D.

\section{Effectively Infinite Systems} \label{Infinity}

The simplified system of evolution considered in the main text assumes that the space of potential population states is infinite, and focuses on unbounded evolution (i.e., $|R_E|=\infty$). One might argue, however, that any real system of evolution is necessarily finite, if only because of a potential limit to the constituent elements of the genetic material. There are two potential responses to this objection. First, on a philosophical level, although any particular evolutionary system might be finite, one might nevertheless want evolutionary theory to stand abstractly, independent of any particular instantiation of an evolutionary dynamic. This is very much analogous to the fact that, in the context of number theory, although one necessarily only ever has to deal with a finite number of things that require counting, we nevertheless desire an abstract theory of numbers that does not presuppose any finite limitations. And just as such a negation-complete theory of numbers is not possible \citep{Godel:1931, Nagel:1958, Davis:1965, Heijenoort:1967, Smith:2007}, neither is one for evolutionary biology unless evolution is progressive.

Second, on a more practical level, it is clear that the digital nature of heredity offered by DNA/RNA makes such systems effectively infinite in that the number of possible population states is enormous. The remainder of this section makes the notation of effectively infinite precise. For simplicity, the focus below is on the deterministic system. 

Recall that, in the $|R_E|=\infty$ case, a function is computable (and total) if it can be evaluated in a finite number of steps, for any input \citep{Cutland:1980} (Appendix \ref{comp}). Thus the predicate `$x \in R_E$' is decidable if its characteristic function can be evaluated, for any input value $x$, in a finite number of steps. Likewise, the mapping $\hat{C}$ of the theorem is computable if, for any input, it returns a code number in a finite number of steps.

When $|R_E|<\infty$, however, the predicate `$x \in R_E$' is always decidable because we can always carry out a complete cataloguing of $R_E$ in a finite number of steps. We simply need to successively evaluate $\phi_E(n)$ for increasing values of $n$. According to Lemma 1 of Appendix \ref{stoch}, because $R_E$ is finite, we will eventually obtain a value that has previously been visited, and from that point onward the system will then simply revisit previously visited states. 

Although these observations are formally correct, they nevertheless fail to capture the important consequences of digital inheritance in finite systems. In particular, the natural analogue of computability for such finite systems in the context of indefinite heredity is not the requirement that an output be obtained in a finite number of steps. Rather, it is that an output be obtained in a finite number of steps, \emph{and that this number of steps not exceed some finite bound that is independent of the size of the state space, $|R_E|$}. For example, with this definition for finite state spaces, the predicate `$x \in R_E$' would be decidable if its characteristic function can be evaluated in a finite number of steps, and if this number never exceeds some finite bound that is independent of $|R_E|$. Thus, regardless of the size of $|R_E|$, we are guaranteed to never need more than a fixed number of computational steps. 

To formalize these ideas, we need to be precise about what it means to consider state spaces of different sizes, $|R_E|$. We do this as follows. First, consider the infinite state space situation used in the main text, where $\phi_E(n)$ denotes the computable function corresponding to the evolutionary process. Next, define the finite state space process by a computable function, $F_E^{\eta}(n)$, where $n=\eta+1$ is the first time at which a previously visited population state is re-visited, and where $F_E^{\eta}(n)=\phi_E(n)$ for all $n \leq \eta$. Note that we have $\eta=|R_E|$, and thus $\eta$ is the state space size. In this way, any given finite state space process is identical to the reference infinite state space process, $\phi_E(n)$, over time until the point $\eta +1$ at which the finite process begins to revisit previously visited states. Thus we can consider state spaces of different sizes, $\eta$, with the limiting case of $\eta \rightarrow \infty$ corresponding to the infinite state space of the main text. We have the following revised definitions for the finite case:

\textbf{Definition}: The predicate `$x \in R_E$' is *decidable if, for any input $x$, there exists a $T<\infty$ such that the characteristic function $c_{R_E}(x)$ can be evaluated in no more than $T$ steps, where $T$ is independent of $\eta$ (i.e., independent of system size). 

\textbf{Definition}: A one-to-one mapping of the population states by the positive integers, $\hat{C}$, is *computable if, for any input there exists a $T<\infty$ such that the mapping can be evaluated in no more than $T$ steps, where $T$ is independent of $\eta$. 

The main theorem of the text can again be seen to hold when $|R_E|<\infty$ if we use the above definitions. In particular, 

\emph{Theorem: `$x \in R_E$' is *decidable if, and only if, there exists an *computable one-to-one coding of the population states by a subset of the positive integers, $\hat{C}$, such that the corresponding description of the evolutionary process, $F^{\eta}_{\hat{E}}(n)$, satisfies $F^{\eta}_{\hat{E}}(n+1) > F^{\eta}_{\hat{E}}(n)$ for all $n \leq \eta$.}

Notice that there is one difference from the main theorem of the text; namely, the altered characterization of progressive evolution. Now, because $R_E$ is finite, we say that evolution is progressive if there is some quantity that increases over time before the process begins to repeat. Also note that, in addition to the altered definition of `computable' and `decidable' in the statement of the theorem, all other instances of computability use this altered definition as well.

Only a sketch of a formal proof is given for this modified theorem because it is similar that of the main text. Recall that $F_E^{\eta}(n)$ denotes the computable function corresponding to the finite evolutionary system of interest.

Proof (Sketch):

Part 1: $\exists$ *$\hat{C} \mbox{ s.t. } F^{\eta}_{\hat{E}}(n+1) > F^{\eta}_{\hat{E}}(n) \mbox{ } \forall n \leq \eta$ $\Rightarrow$ `$x \in R_E$' *decidable

As before, take any input $x$ and find its new code, $\hat{x}$. By hypothesis the number of steps required is bounded by a constant that is independent of system size. Next, we can begin to successively evaluate $F^{\eta}_E(n)$ for increasing values of $n$. We suppose that the number of steps required in this computation for any $n \leq \eta$ is independent of $\eta$. This is a reasonable assumption because the outputs are identical to those of $\phi_E(n)$ when $n \leq \eta$, and the number of steps required to evaluate $\phi_E(n)$ is independent of $\eta$ for any $n$. To each output of $F^{\eta}_E(n)$ we can apply the above mapping, $\hat{C}$ to obtain $F^{\eta}_{\hat{E}}(n)$, which by hypothesis, increases with $n \leq \eta$. By hypothesis the number of steps required is independent of $\eta$ for each such application.

As we proceed, either we reach (i) $n=\eta$ prior to reaching an $n$ for which $\hat{x}<F^{\eta}_{\hat{E}}(n)$, or we reach (ii) a value of $n$ whereby $\hat{x}<F^{\eta}_{\hat{E}}(n)$ before $n=\eta$. In either case `$x \in R_{\hat{E}}$' is then decidable because, if $\hat{x}$ has not been reached by this point, it never will be. Thus, `$x \in R_E$' is decidable as well. Moreover, if (i) pertains, then the number of steps required before deciding is no more than $\mu i (\phi_{\hat{E}}(i) \geq \hat{x})$, If (ii) pertains, then this number of steps is exactly equal to $\mu i (\phi_{\hat{E}}(i) \geq \hat{x})$. And because $\mu i (\phi_{\hat{E}}(i) \geq \hat{x})$ is finite and independent of $\eta$, we can see that `$x \in R_E$' is *decidable as well.

Part 2: `$x \in R_E$' *decidable $\Rightarrow$ $\exists$ *$\hat{C} \mbox{ s.t. } F^{\eta}_{\hat{E}}(n+1) > F^{\eta}_{\hat{E}}(n) \mbox{ } \forall n \leq\eta$

We can construct the required *computable bijection between population states and an appropriate coding as follows. First, take any effective coding of population states. By hypothesis, the number of steps required to decide `$x \in R_E$' for any $x$ is finite and independent of $\eta$. Thus, we can proceed through the population states, $x$, in increasing order, applying the following algorithm:

(i) if $x \notin R_E$ and it is the $k^{th}$ such state up to that point, use the $k^{th}$ odd number as its new code.

(ii) if $x \in R_E$, calculate $\mu i (F^{\eta}_E(i)=x)$, and use the $i^{th}$ even number as its new code.

As we proceed though the states, $x$, the number of steps required for each, regardless of whether (i) and (ii) pertains, is independent of $\eta$. Therefore, the entire coding procedure for any given state is independent of $\eta$ as well; i.e., the coding is *computable as required.

\end{document}